\documentclass{jetpl}
\twocolumn

\usepackage{amsmath,amssymb,epsfig,color,pstricks,bm,graphics,alltt}

\usepackage[colorlinks,plainpages=false,linkcolor=black,urlcolor=blue,citecolor=black,pdfpagemode=UseNone,pdfstartview=FitBH]{hyperref}

\definecolor{MyDarkBlue}{rgb}{0,  0.3,  0.9}
\definecolor{MyDarkBlack}{rgb}{0,  0,  0}
\newcommand \modified[1]{\textcolor{black}{#1}}

\begin{document}

\lat

\title{Consistent LDA$^{\prime}$+DMFT -- an unambiguous way to avoid double counting problem: NiO test}


\rtitle{Consistent LDA$^{\prime}$+DMFT}

\sodtitle{Consistent LDA+DMFT}

\author{$^{a}$I.\ A.\ Nekrasov\thanks{E-mail: nekrasov@iep.uran.ru},
$^{a}$N.\ S.\ Pavlov\thanks{E-mail: pavlov@iep.uran.ru}, 
$^{a,b}$M.\ V.\ Sadovskii\thanks{E-mail: sadovski@iep.uran.ru}}

\rauthor{I.\ A.\ Nekrasov, N.\ S.\ Pavlov, M.\ V.\ Sadovskii}

\sodauthor{Nekrasov, Pavlov, Sadovskii}

\sodauthor{Nekrasov, Pavlov, Sadovskii}

\address{$^a$Institute for Electrophysics, Russian Academy of Sciences, 
Ural Branch, Amundsen str. 106,  Ekaterinburg, 620016, Russia\\
$^b$Institute for Metal Physics, Russian Academy of Sciences, Ural Branch,
S.Kovalevskoi str. 18, Ekaterinburg, 620990, Russia}

\dates{March 2012}{*}

\abstract{
We present a consistent way of treating a double counting 
problem unavoidably arising within the LDA+DMFT combined approach to
realistic calculations of electronic structure of strongly correlated systems.
The main obstacle here is the absence of systematic (e.g. diagrammatic) way to 
express LDA (local density approximation) contribution to exchange correlation 
energy appearing in the density functional theory. It is not clear then, 
which part of interaction entering DMFT (dynamical mean-field theory) 
is already taken into account through LDA calculations. Because of that, up to 
now there is no accepted unique expression for the double counting correction
in LDA+DMFT. To avoid this problem we propose here the consistent 
LDA$^{\prime}$+DMFT approach, where LDA exchange correlation contribution is 
explicitly excluded for correlated states (bands) during self-consistent
band structure calculations. What is left out of Coulomb interaction for those 
strongly correlated states (bands) is its non-local part, which is not included 
in DMFT, and the local Hartree like contribution. Then the double counting 
correction is uniquely reduced to the local Hartree contribution. 
Correlations for strongly correlated states are then directly accounted for via 
the standard DMFT. We further test the consistent LDA$^{\prime}$+DMFT scheme 
and compare it with conventional LDA+DMFT calculating the electronic structure 
of NiO. Opposite to the conventional LDA+DMFT our consistent 
LDA$^{\prime}$+DMFT approach unambiguously produces the insulating band 
structure in agreement with experiments.
}

\PACS{71.20.-b, 71.27.+a, 71.28.+d, 74.25.Jb, }
                          
\maketitle

\section{Introduction}

During the last 15 years the so called LDA+DMFT approach
(local density approximation + dynamical mean-field theory) became
a common tool to describe band structure of real strongly correlated
materials \cite{poter97,LDADMFT1,Nekrasov00, psik, LDADMFT,IzAn}.
In this approach the results of LDA band structure calculations are
supplemented with local Coulomb (Hubbard) interaction term
for those states which are counted as strongly correlated.
Formally the LDA+DMFT Hamiltonian can be written as
\begin{eqnarray} \hat{H} &=&
\hat{H}_{\rm LDA} -\hat{H}^{DC}+
 \frac{1}{2} {\sum_{i=i_d, l= l_d}
\sum_{m\sigma, m^{\prime }\sigma'}
}^{\!\!\!\!\!\prime } \;\; 
{U_{mm^{\prime}}^{\sigma \sigma'}}\hat{n}_{ilm\sigma}\hat{n}_{ilm'\sigma'}
\nonumber \\&&
-\frac{1}{2} {\sum_{i=i_d, l= l_d}}
\sum_{m\sigma, m^{\prime }\bar\sigma}^{\;\;\;\;\;\;\;\;\;\prime}
{J_{mm'}^{}}
\hat{c}^\dagger_{ilm\sigma }
\hat{c}^\dagger_{ilm^{\prime}\bar{\sigma}} 
\hat{c}^{\phantom{\dagger}}_{ilm^{\prime}\sigma} 
\hat{c}^{\phantom{\dagger}}_{ilm\bar{\sigma}}.
\label{Hint}
\end{eqnarray}
Here $U_{mm^{\prime }}^{\sigma
\sigma ^{\prime }}$ are the most important matrix elements of Coulomb matrix
(Coulomb repulsion and $z$-component of
Hund's rule coupling) and ${J_{mm'}}$ are spin-flip terms
of Hund's rule couplings between the strongly correlated electrons (assumed here
to be $d$-states, enumerated by $i=i_{d}$ and $l=l_{d}$).
The prime on the sum indicates that at least two of the indices of operators 
have to be different, and
$\bar{\sigma}= \downarrow\!(\uparrow)$ for $\sigma = \uparrow\!(\downarrow)$. 

The LDA part of the Hamiltonian (\ref{Hint}) is given by:
\begin{eqnarray}
\hat{H}_{{\rm LDA}} &=& -\frac{\hbar ^{2}}{2m_{e}}\Delta +V_{{\rm ion}}({\bf r})
+\int d^3{r^{\prime }}\,\rho ({\bf r^\prime })V_{ee}({\bf r}\!-\!{\bf r^\prime})\nonumber \\
&+&\frac{\delta E_{\rm xc}^{\rm LDA}(\rho )}{\delta \rho({\bf r})}
\label{HLDA0},
\end{eqnarray}%
where $\Delta $ is the Laplace operator, $m_{e}$ the electron mass, $e$ the electron
charge, and
\begin{eqnarray}
V_{{\rm ion}} ({\bf r})=-e^2\sum_i \frac{Z_i}{|{\bf r}-{\bf R_i}|},
&\;\; & V_{\rm ee}({\bf r}\!-\!{\bf r'})=\frac{e^2}{2}
\sum_{{\bf r} \neq {\bf r'}} \frac{1}{|{\bf r}-{\bf r'}|}\nonumber\\
\end{eqnarray}
denote the one-particle potential due to all ions $i$
with charge $eZ_{i}$ at given positions ${\bf R_{i}}$, and the
electron-electron interaction, respectively.

The $E_{\rm xc}^{\rm LDA}(\rho ({\bf r}))$ in (\ref{HLDA0}) is a function of
local charge density which approximates true exchange correlation
functional $E_{\rm xc}[\rho ]$ of density functional theory in the framework of 
local density approximation \cite{JonesGunn}.
The form of the function $E_{\rm xc}^{\rm LDA}(\rho ({\bf r}))$
is usually calculated from perturbation theory \cite{jellium}
or numerical simulations \cite{jellium2} of the ``jellium'' model with 
$V_{\rm ion}({\bf r})={\rm const}$.
Once we choose some basis set of one-particle wave functions $\varphi _{i}$
(e.g. to do practical calculations and explicitly express matrix elements of the
Hamiltonian (\ref{HLDA0})), we can obtain $\rho $ as:
\begin{equation}
\rho ({\bf r})=\sum_{i=1}^{N}|\varphi _{i}({\bf r})|^{2}.
\label{rhophi}
\end{equation}%
Finally a term $\hat{H}^{DC}$ is subtracted in Eq. (\ref{Hint}) to avoid
double-counting of those contributions of the {\em local} Coulomb
interaction already contained in $\hat{H}_{{\rm LDA}}$ via Hartree term and
$E_{\rm xc}^{\rm LDA}(\rho ({\bf r}))$.
Since there does not exist a direct microscopic or diagrammatic link
between the model (Hubbard like) Hamiltonian approach and LDA it is not possible 
to express $\hat{H}^{DC}$ rigorously in terms of $U$, $J$ and $\rho$. 
Thus there is no unique and accepted expression for
$\hat{H}^{DC}$ (see e.g. Ref.~\cite{Karolak}).

One popular expression for $\hat{H}^{DC}$ is the Hartree like 
(fully localized limit) expression \cite{Anisimov91}:
\begin{equation}
H^{DC}=\frac{1}{2}U n_{d}(n_{d}-1)-
\frac{1}{2}{J}\sum_\sigma n_{d\sigma}(n_{d\sigma}-1).
\label{ELDAU}
\end{equation}%
Here, $n_{d\sigma}=\sum_{m}n_{il_{d}m\sigma}=\sum_{m}\langle \hat{n}%
_{il_{d}m\sigma}\rangle $ is total number of
electrons on interacting orbitals per spin, $n_d=\sum_\sigma n_{d\sigma}$,
$U$ is Coulomb (Hubbard) repulsion and $J$ is the exchange
or Hund's rule coupling obtained usually from constrained
LDA procedure \cite{Gunnarsson}.
The  $n_d$ value can be obtained either from LDA calculations
or can be recalculated during the DMFT loop. Practically, the values obtained 
are pretty close to each other.

Below we introduce the consistent LDA$^{\prime}$+DMFT approach,
which allows one to avoid the double counting problem unambiguously.
To illustrate the advantages of this new approach we shall apply it to
calculations of the band structure of the well known prototype of charge 
transfer insulating system NiO.

\section{Consistent LDA$^{\prime}$+DMFT approach}

One of the possible ways to solve the double counting problem is
to perform Hartree+DMFT or Hartree-Fock+DMFT calculations
(see for the overview of the concept Ref.~\cite{Held}).
This approach uses the advantage of knowledge of diagrammatic expression for
Hartree or Hartree-Fock terms. Thus, performing 
Hartree-Fock band structure calculations for real materials
we do know exactly what portion of interaction is, in fact, explicitly included.
Then obviously, the double counting term should be chosen in the form of
Eq.~(\ref{ELDAU}). However, up to now we are unaware of any Hartree+DMFT
or Hartree-Fock+DMFT calculations for real materials.

In fact, Hartree-Fock band structure calculations are in some sense a large
step backwards from DFT/LDA approach, which was so successful in description of many 
real materials. Even in the case of strongly correlated systems DFT/LDA is
recognized as a best starting point for further model Hamiltonian treatments,
such as e.g. LDA+DMFT method.

In view of this we suggest a kind of compromise between Hartree-Fock and DFT/LDA 
starting points to be followed by DMFT calculations.
As described above main obstacle to express double counting term
exactly is exchange correlation $E_{\rm xc}^{\rm LDA}(\rho ({\bf r}))$
portion of interaction within LDA. It seems somehow inconsistent to use it to 
describe correlation effects in narrow (strongly correlated) bands from the very
beginning, as these should be treated via more elaborate schemes like DMFT.  
To overcome this difficulty for these states, we propose 
to redefine charge density (\ref{rhophi}) in $E_{\rm xc}^{LDA}$ as follows:
\begin{equation}
\rho^\prime ({\bf r})=\sum_{i\neq i_d}|\varphi _{i}({\bf r})|^{2}
\label{rhophi1}
\end{equation}
{\em excluding the contribution of the density of strongly correlated electrons}.
Then this redefined $\rho^\prime ({\bf r})$ is used to obtain 
$E_{\rm xc}^{\rm LDA}$ and perform the self-consistent LDA band structure 
calculations for correlated bands at the initial stage of LDA+DMFT, while 
correlations of $d$-electrons are left to be treated via DMFT. This means that 
what is left for correlated states out of interaction on the LDA stage would be 
just the Hartree contribution of Eq.~(\ref{HLDA0}). At the same time all other 
states (not counted as strongly correlated) are to be treated with the full 
power of DFT/LDA and {\em full} $\rho$ in $E_{\rm xc}^{LDA}$. Now, the problem 
of double counting correction is uniquely defined -- it should be taken in  the 
form of the Hartree like term, given by  Eq.~(\ref{ELDAU}). 

This approach to describe realistic strongly correlated
systems we shall call the consistent LDA$^{\prime}$+DMFT.
It is in precise correspondence with the standard definition of
correlations, as interaction corrections ``above'' Hartree-Fock. We explicitly
exclude contributions to $E_{\rm xc}^{LDA}$ from (strongly) correlated bands,
where correlations are treated via DMFT, while we take all electrons into 
account in LDA calculations for all other (non correlated) bands.

\section{Results}

Following many recent works \cite{Karolak,Ren,Kunes} (and references therein)
we choose as a testing system the prototype charge transfer insulator NiO.
LDA band structure calculations for NiO were performed
within the linearized muffin-tin orbitals (LMTO) basis set \cite{LMTO}.
In the corresponding program package TB-LMTO v.47 the $E_{\rm xc}^{\rm LDA}$
was taken in von Barth-Hedin form \cite{jellium}.

\begin{figure*}
\includegraphics[clip=true,width=0.75\textwidth]{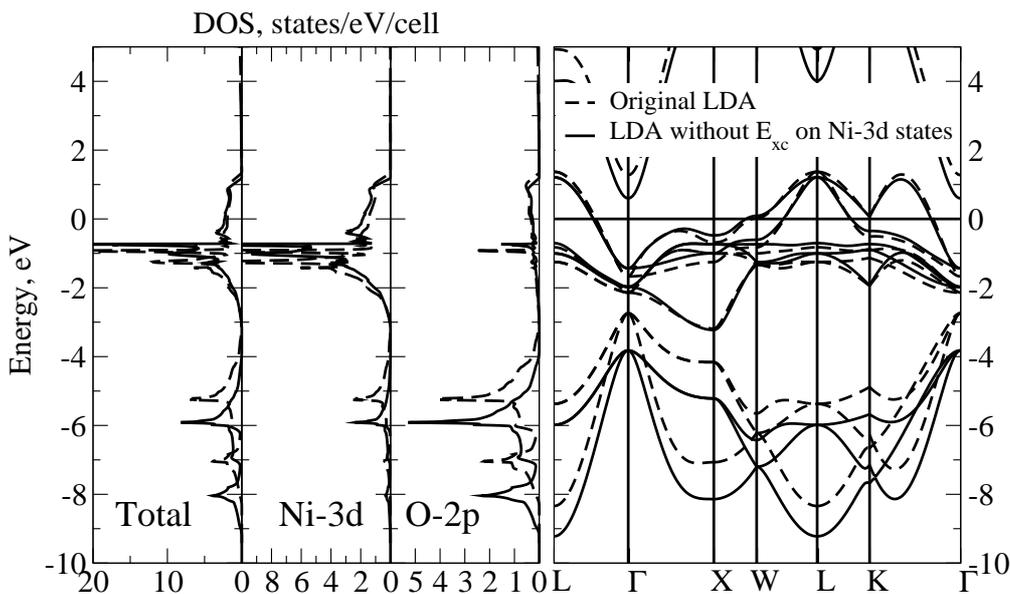}
\caption{Fig. 1. LDA and LDA$^{\prime}$ calculated band dispersions and densities 
of states of NiO. The Fermi level $E_F$ is at zero energy.} 
\end{figure*}

In the Fig.~1 we present LDA densities of states (left panel) and band dispersions
(right panel) of NiO. Band dispersions consist of two separate sets
of bands: the O-2p bands (from -3 to -9 eV) and Ni-3d bands, crossing the Fermi level 
(from 1.5 to -3 eV). Dashed lines in Fig.~1 show conventional LDA results. 
Full lines correspond to LDA$^\prime$ calculations without $E_{\rm xc}^{\rm LDA}$ on 
Ni-3d states, namely, with redefined charge density (\ref{rhophi1}) in 
$E_{\rm xc}^{\rm LDA}$. Overall changes can be characterized as
an almost rigid shift of oxygen states down in energy by about 1 eV for
LDA$^\prime$ calculations, while Ni-3d states are only slightly modified
due to change of Ni-O hybridization. In other words LDA$^\prime$ calculations lead 
to the change of charge transfer energy $\Delta=|\varepsilon_d-\varepsilon_p|$ by 
about 1 eV. Rather small influence of $E_{\rm xc}^{\rm LDA}$ on Ni-3d states 
is not surprising, since $E_{\rm xc}^{\rm LDA}$ for metallic (LDA produces 
metallic state for NiO) electron densities $r_s$=2--6 are known to be of the 
order of 1 eV \cite{jellium2}.
Further we perform DMFT calculations using LDA and LDA$^\prime$ Hamiltonians,
which include all states (without any projecting). DMFT impurity solver used
was Hirsh-Fye quantum Monte-Carlo algorithm \cite{QMC}.
Inverse temperature was taken $\beta=5$eV$^{-1}$ (2321~K) and 80 time slices were
used, with 10$^6$ Monte Carlo sweeps. The use of very high temperature does not
lead to any qualitative effects in the results, allowing us to avoid 
unnecessary computational efforts. Parameters of Coulomb interaction were chosen as 
typical for NiO \cite{Karolak,Kunes}: $U$=8~eV and $J$=1~eV. To obtain DMFT(QMC) densities 
of states at real energies, we employed the maximum entropy method \cite{MEM}.

In the Fig.~2 we compare the conventional LDA+DMFT (upper panel) and consistent
LDA$^{\prime}$+DMFT (lower panel) results for NiO. Different lines represent
partial Ni-3d($t_{2g}$) (solid line), Ni-3d($e_g$) and oxygen O-2p (dash-dot line)
contributions to density of states. To obtain O-2p states DMFT(QMC) self-energy was 
analytically continued to real frequencies by Pade approximant method.
For both conventional LDA+DMFT and consistent LDA$^{\prime}$+DMFT calculations
we used $H^{DC}$ of Eq. (\ref{ELDAU}) with $n_d$ recalculated on each DMFT iteration
step. Corresponding values of $H^{DC}$ are 62 eV ($n_d$=8.7) and 58.13 eV ($n_d$=8.2)
for conventional LDA+DMFT and consistent LDA$^{\prime}$+DMFT respectively.
The total occupancies of Ni-3d states within LDA and LDA$^\prime$ calculations
were 8.5 and 8.3.

Within conventional LDA+DMFT we obtain the metallic solution, which contradicts
experiments. This fact can be explained as follows. We already mentioned that
LDA and LDA$^\prime$ calculations results differ mainly by the values of
charge transfer energy $\Delta=|\varepsilon_d-\varepsilon_p|$. In fact, 
we observed \cite{Pavlov} that double counting correction essentially affects $\Delta$,
or the other way around, the different values of $\Delta$ require the different values of 
double counting corrections to obtain the same results. In its turn, the different values 
of double counting correction can lead either to metallic or insulating solutions for the 
same set of other parameters \cite{Karolak,Pavlov}.

Once we employ the consistent LDA$^{\prime}$+DMFT approach, we obtain the
charge transfer {\em insulating} solution for NiO, which agrees well
with other LDA+DMFT calculations for NiO \cite{Karolak,Kunes}
and experiment \cite{Sawatzky}, confirming the effectiveness of our
approach. Namely, the peak at -2 eV which consists almost
in equal parts from Ni-3d and O-2p states is nothing else but
Zhang-Rice bound state (in agreement with Ref.~\cite{Kunes}).
Lower Hubbard band formed mainly from Ni-3d states is located
lower in energy than Zhang-Rice band. Conducting
band is just the upper Hubbard band dominated by Ni-3d states.

As an additional check of consistency of our approach we also performed 
LDA$^{\prime}$+DMFT calculations for SrVO$_3$. The results obtained are in 
good agreement with those obtained in Ref.~\cite{Leherman}, further validating 
our proposed LDA$^{\prime}$+DMFT approach as an effective and unambiguous method
of band structure calculations for strongly correlated systems.

\begin{figure}
\includegraphics[clip=true,width=0.45\textwidth]{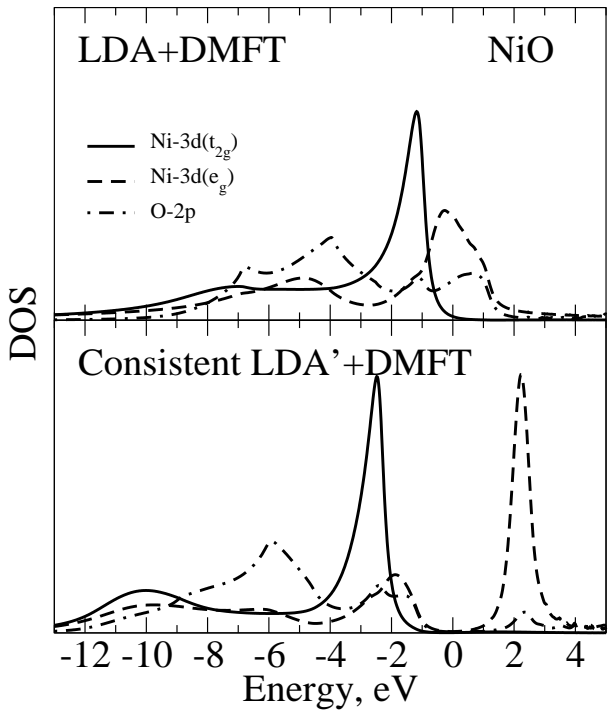}
\caption{Fig. 2. Consistent LDA$^{\prime}$+DMFT (lower panel) and LDA+DMFT (upper panel)
partial densities of states for NiO. The Fermi level is at zero energy.} 
\end{figure}

\section {Conclusion}

In this work we proposed the consistent LDA$^{\prime}$+DMFT approach, which 
solves the problem of non-uniqueness of the double counting correction.
By excluding LDA exchange correlation contribution for correlated states
within the self-consistent LDA calculations
(e.g. for Ni-3d states) we end up with just Hartree like portion of interaction 
for (strongly) correlated states. Then we know exactly, what should be 
subtracted as a double counting correction term, while merging LDA$^\prime$ and 
DMFT. We tested our consistent LDA$^{\prime}$+DMFT approach, calculating the 
band structure of NiO. We obtained the insulating solution without any 
additional fitting parameters and in general agreement with experimental 
data \cite{Sawatzky}, while in other LDA+DMFT works for NiO the double counting 
correction was either treated as an adjustable parameter \cite{Karolak}, or the
special form of double counting term was introduced \cite{Kunes} to achieve
agreement with experiment.

We thank A. Poteryaev for providing us QMC code and many helpful discussions.
This work is partly supported by RFBR grant 11-02-00147 and was performed
within the framework of programs of fundamental research of the Russian 
Academy of Sciences (RAS) ``Quantum mesoscopic and disordered structures'' 
(12-$\Pi$-2-1002) and of the Physics Division of RAS  ``Strongly correlated 
electrons in solids and structures'' (012-T-2-1001).


\begin{thebibliography}{99}

\bibitem{poter97}  V.I. Anisimov, A.I. Poteryaev, M.A. Korotin, A.O. Anokhin and G. Kotliar, J. Phys. Cond. Matter {\bf 9}, 7359 (1997).

\bibitem{LDADMFT1}
A.I.  Lichtenstein, M.I. Katsnelson, Phys. Rev. B {\bf 57}, 6884 (1998).

\bibitem{Nekrasov00}  I.A. Nekrasov, K. Held, N. Bl\"umer, A.I. Poteryaev, V.I. Anisimov and D. Vollhardt, Euro. Phys. J. B {\bf 18}, 55 (2000).

\bibitem{psik} K. Held, I.A. Nekrasov, G. Keller, V. Eyert, N. Bl\"umer, A.K. McMahan, R.T. Scalettar, T. Pruschke, V.I. Anisimov, D. Vollhardt, Psi-k Newsletter {\bf 56}, 65 (2003).

\bibitem{LDADMFT} K. Held, I.A. Nekrasov, N. Bl\"umer, V.I. Anisimov and D. Vollhardt, Int. J. Mod. Phys. B {\bf 15}, 2611 (2001); 
K. Held, I.A. Nekrasov, G. Keller, V. Eyert, N. Bl\"umer, A.K. McMahan, R.T. Scalettar, T. Pruschke, V.I. Anisimov, and D. Vollhardt in {\it Quantum Simulations of
Complex Many-Body Systems: From Theory to Algorithms}, (Eds. J. Grotendorst, D. Marks, A. Muramatsu) NIC Series Volume 10 (2002) p. 175;
A.I. Lichtenstein, M.I. Katsnelson, G. Kotliar G, in {\em Electron Correlations and Materials Properties 2nd ed.} (Eds. A. Gonis, N. Kioussis, M. Ciftan) (Kluwer Academic/Plenum, 2002) p. 428.

\bibitem{IzAn} V.I. Anisimov, Yu.A. Izyumov,  {\it Electronic Structure of Strongly Correlated
Materials }(Berlin - Heidelberg: Springer, 2010 ).

\bibitem{JonesGunn}  R. O. Jones and O. Gunnarsson, Rev. Mod. Phys. {\bf 61},
689 (1989).

\bibitem{jellium}
L. Hedin and B. Lundqvist, J. Phys. C: Solid State Phys. \textbf{4}, 2064 (1971);
U. von Barth and L. Hedin, J. Phys. C: Solid State Phys. \textbf{5}, 1629 (1972).

\bibitem{jellium2}
D.~M. Ceperley and B.~J. Alder, Phys. Rev. Lett. 45, 566 (1980).

\bibitem{Karolak} M. Karolak, G. Ulm, T. Wehling, V. Mazurenko, A. Poteryaev, A. Lichtenstein, Journal of Electron Spectroscopy and Related Phenomena, Volume {\bf 181}, 11 (2010).

\bibitem{Anisimov91}  V. I. Anisimov, J. Zaanen, and O. K. Andersen, Phys.
Rev. B {\bf 44}, 943 (1991); V.~I.~Anisimov, F. Aryasetiawan, and A. I.
Lichtenstein, J. Phys. Cond. Matter {\bf 9}, 767 (1997).

\bibitem{Gunnarsson} O.\,Gunnarsson, O.\,K.\,Andersen, O.\,Jepsen, and J.\,Zaanen,
Phys. Rev. B \textbf{39}, 1708 (1989).

\bibitem{Held} K. Held, Advances in Physics {\bf 56}, 829 (2007). (see page 862)

\bibitem{Ren} Ren X., Leonov I., Keller G., Kollar M., Nekrasov I., Vollhardt D., Phys. Rev. B. {\bf 74}, 195114 (2006).

\bibitem{Kunes}J. Kunes, V. I. Anisimov, S. L. Skornyakov, A. V. Lukoyanov, and D. Vollhardt, Phys. Rev. Lett. \textbf{99}, 156404 (2007);
J. Kunes, V. I. Anisimov, A. V. Lukoyanov, and D. Vollhardt, Phys. Rev. B. \textbf{75}, 165115 (2007).

\bibitem{LMTO} O. K. \modified{Andersen} , Phys. Rev. B {\bf 12}, 3060 (1975);
O. K. Andersen and O. Jepsen, Phys. Rev. Lett. {\bf 53}, 2571 (1984).

\bibitem{QMC}  J. E. Hirsch and R. M. Fye, Phys. Rev. Lett. {\bf 56}, 2521
(1986); M. Jarrell, Phys. Rev. Lett. {\bf 69}, 168 (1992); M. Rozenberg,
X. Y. Zhang, and G. Kotliar, Phys. Rev. Lett. {\bf 69}, 1236 (1992);
A. Georges and W. Krauth, Phys. Rev. Lett. {\bf 69}, 1240 (1992);
M.~Jarrell, in {\it
Numerical Methods for Lattice Quantum Many-Body Problems}, edited by
D.~Scalapino, Addison Wesley, 1997.

\bibitem{MEM}
M.~Jarrell and J.~E. Gubernatis,
\newblock { Physics Reports \bf 269}, 133 (1996).

\bibitem{Pavlov} N.S. Pavlov, I.A. Nekrasov, E.Z. Kuchinskii, to be published.

\bibitem{Sawatzky} G.A. Sawatzky and J.W. Allen, Phys. Rev. Lett. {\bf 53}, 2339 (1984).

\bibitem{Leherman} B. Amadon, F. Lechermann, A. Georges, F. Jollet, T.O. Wehling, A.I. Lichtenstein, Phys. Rev. B {\bf 77}, 205112 (2008).

\end{thebibliography}
\end{document}